\documentclass{ws-procs9x6-cpt16}
\begin{document}

\newcommand{\refeq}[1]{(\ref{#1})}
\def\etal {{\it et al.}}

\title{Acoustic Tests of Lorentz Symmetry using\\
Bulk Acoustic Wave Quartz Oscillators}

\author{M.\ Goryachev,$^1$ A.\ Lo,$^{2}$ Ph.\ Haslinger,$^2$
E.\ Mizrachi,$^2$ L.\ Anderegg,$^2$\\
H.\ M\"uller,$^2$ M.\ Hohensee,$^{2,3}$ and M.E.\ Tobar$^1$}

\address{$^1$ARC Centre of Excellence for Engineered Quantum Systems, 
School of Physics\\
University of Western Australia, 35 Stirling Highway, Crawley WA 6009, Australia}
\address{$^2$Department of Physics, University of California,
Berkeley, CA 94720, USA}

\address{$^3$Lawrence Livermore National Laboratory, Livermore, CA 94550, USA}

\begin{abstract}
A new method of probing Lorentz invariance in the neutron sector is described. The method is baed on stable quartz bulk acoustic wave oscillators compared on a rotating table. Due to Lorentz-invariance violation, the resonance frequencies of acoustic wave resonators depend on the direction in space via a corresponding dependence of masses of the constituent elements of solids. This dependence is measured via observation of oscillator phase noise built around such devices. The first such experiment now shows sensitivity to violation down to the limit $\tilde{c}^n_Q=(-1.8\pm2.2)\times 10^{-14}$ GeV. Methods to improve the sensitivity are described together with some other applications of the technology in tests of fundamental physics. 
\end{abstract}

\bodymatter

\phantom{}\vskip10pt\noindent
The possibility that physics beyond the Standard Model might violate Lorentz invariance\cite{ColladayKostelecky,KosteleckyMewes,KosteleckyMewesPRD} has motivated a broad range of precision tests of fundamental properties of Nature. Typically in such experimental tests, one compares two clocks that probe two orthogonal directions in space and measure frequency deviations due to Lorentz-invariance violations over significant amount of time. This time is required to collect enough statistics associated with clock rotations and boosts. Usually, three types of clock rotation in the Sun-centred frame are taken into account: Earth rotation around the Sun, Earth rotation around its axis, and rotation of an experimental setup on a turntable. Considering these requirements, many tests of Lorentz symmetry are not limited by clock stability but often by systematic effects from wobble and tilt of the turntable, and the ability to acquire data over long stretches of time. 
Due to these difficulties, not all sectors of the Standard-Model Extension (photon, neutron, electron, etc.) are explored equally well. Moreover, within one sector some coefficients might be very well bounded by, for instance, using astrophysical observations and others are left untouched. Thus, there is a need for novel approaches to test Lorentz symmetries in all sectors as well as to cross check techniques. 

A new method of studying Lorentz invariance in the neutron sector using quartz bulk acoustic wave (BAW) oscillators and resonators has been recently proposed.\cite{PhRevX} This approach is based on the fact that violations of Lorentz invariance in the matter sector generate anisotropies in the inertial masses of particles and the elastic constants of solids, giving rise to measurable anisotropies in the resonance frequencies of acoustic modes in mechanical resonators. It means that mechanical properties of solids and thus associated acoustical frequencies of BAW resonators depend on the direction in space via the Lorentz anisotropy of particles.
While nowadays the frequency stability of quartz BAW oscillators is surpassed by atomic clocks, it offers a simple, robust and reliable method of time keeping with oscillating masses with all systematics well understood and characterized. Moreover, quartz BAW oscillators are the most stable frequency references based on the mechanical motion of macroscopic objects.

\begin{figure}
\begin{center}
\includegraphics[width=4in]{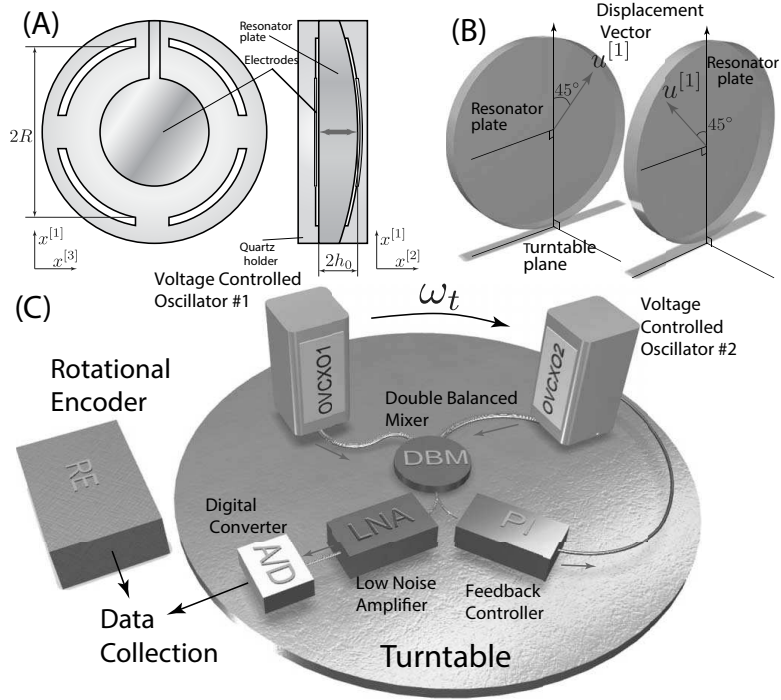}
\end{center}
\caption{(A) BAW quartz resonator, top and cut views. (B) Orientation of the displacement vectors for two resonators on the turntable. (C) Rotating experimental setup. Additional magnetic shielding is not shown.}
\label{goryachevfig1}
\end{figure}

Two room-temperature voltage-controlled quartz oscillators ($2\times10^{-12}$ frequency stability) are used in the current version of the experiment. These oscillators are based on stress-compensated cut quartz BAW resonators working on the slow shear thickness mode exhibiting a frequency-temperature turnover point (Fig.~\ref{goryachevfig1}A). The oscillators are placed inside magnetic shields on a turntable in such a way that the corresponding displacement vectors point in orthogonal directions (Fig.~\ref{goryachevfig1}B). The setup is rotated to induce possible modulation signals associated with the Lorentz anisotropies 
(Fig.~\ref{goryachevfig1}C). In other words, the experiment is a measurement of frequency fluctuations coming from mass variation due to rotation of the experimental setup. Similar setups are used to characterize frequency stability of different types of oscillators. The setup proved to be reliable and robust with all systematics including ageing, vibration, and temperature sensitivity well understood. 

The data have been collected for 120 hours providing 
the frequency resolution of\cite{PhRevX} $2.4\times 10^{-15}$.
The data is analysed in frequency and time domains for both quadratures of the expected signal. No signs of such signals are found at the achieved level of sensitivity. This fact allows us to put a limit of $\tilde{c}^n_Q=(-1.8\pm2.2)\times 10^{-14}$ GeV on the most weakly constrained neutron-sector $c$-coefficient of the Standard-Model Extension, i.e., to rule out all possibilities for Lorentz-violating anisotropies in the inertial masses of neutrons, protons, and electron. This result is found to be a few orders of magnitude improvement over previous laboratory tests and astrophysical bounds.\cite{datatables}

The presented results can be improved further using one of the following approaches: (1) more stable room temperature quartz oscillators (the best frequency stability of room temperature oscillators achieves $2.5\times10^{-14}$);\cite{stability} (2) other methods to probe phonon systems (optomechanics, phonon laser); (3) oscillator arrays (employ crosscorelation techniques to reduce uncorrelated noise that can help to go beyond the single oscillator noise limit); (4) data from BAW quartz oscillators in space missions (DORIS system, satellites Jason-1,2,3);\cite{jason} and (5) cryogenic quartz oscillators (four orders of magnitude improvement in $Q$ factors).\cite{advance} The latter approach is based on the ultra-high quality factors of phonon trapping acoustic resonators 
(approaching $10^{10}$),\cite{gorprl,screp} 
promising frequency stabilities as low as $2\times 10^{-16}$ and, thus, about four orders of magnitude improvement over the current experiment. 

In addition to the described tests of Lorentz invariance in the matter sector, the same BAW quartz technology can be used for other experiments in fundamental physics, such as high-frequency gravity-wave detection\cite{gw} from different proposed sources, dark-matter searches,\cite{dark} probing Planck-scale physics,\cite{plank} etc., as well as in applications of quantum technology.\cite{qubit}

\section*{Acknowledgments}

This work was supported by the Australian Research Council Grant No.\ CE110001013 and DP160100253 as well as the Austrian Science Fund (FWF) J3680.

\end{document}